# Thermal noise engines

Laszlo B. Kish [a]

Texas A&M University, Department of Electrical Engineering, College Station, TX 77843-3128,

USA;  email: Laszlo.Kish@ece.tamu.edu

**Abstract.** In this paper, electrical heat engines driven by the Johnson-Nyquist noise of resistors are introduced. They utilize Coulomb's law and the fluctuation-dissipation theorem of statistical physics (which is the reverse phenomenon of heat dissipation in a resistor). In these engines, resistors, capacitors and switches are the building elements. For best performance, a large number of parallel engines must be integrated to run in a synchronized fashion, and the size of an elementary engine must be at the 10 nanometers scale. At room temperature, in the most idealistic case, a two-dimensional ensemble of engines of 25 nanometer size integrated on a 2.5x2.5 cm silicon wafer with 10 $^o$C temperature difference between the warm-source and the cold-sink would produce a power of about 0.5 Watt. Regular and coherent (correlated-cylinder states) versions of these engines are shown and both of them can operate in either four-stroke or two-stroke modes. In the idealistic case, all these engines have Carnot efficiency, which is the highest efficiency possible in any heat engine without violating the second law of thermodynamics.

engines.

**1. Introduction: quantum leaps and noise bounds**

Energy harvesting applications are aiming the utilization of various forms of energy that would be wasted such as spontaneous thermal gradients. The goal of this paper is to show that the unavoidable nuisance, the thermal noise (Johnson-Nyquist noise), can be put to work and it can drive a "clean", purely electrical heat engine whenever thermal gradient is present.

Additionally, it is interesting to note that John Johnson and Harry Nyquist of Bell Labs discovered/explained the thermal noise voltage of resistors (which is a classical statistical physical phenomenon) [1,2] several years after the completion of the foundations of quantum physics [3]. Similarly, recent applications of the thermal noise for unconventional informatics [4-6], potential competitors of quantum informatics, have emerged many years after the introduction of the corresponding quantum schemes. In the present article, another example is shown where thermal noise-based solutions have claims similar to quantum schemes: heat engines. Some of the thermal noise engines, while they are classical physical objects, show similar properties to coherent quantum heat engines introduced during the last decade [7-11].

---
[a] Until 1999: L.B. Kiss

After general considerations and the investigation of feasibility by the laws of physics, the new type of heat engine schemes will be shown and analyzed.

## 2. General considerations and assumptions

### *2.1 The source of the work in a thermal noise engine*

The "steam" in this heat engine is the Johnson-Nyquist noise, which is a stochastic voltage fluctuation, that appears between the terminals of an open-ended resistor. It is a manifestation of the thermal motion of electrons (and/or holes) in the resistor and it is mathematically described by the fluctuation-dissipation theorem [2,12]. When a resistor is shunted by a capacitor, the capacitor will be charged (thermalized) within the relaxation time constant $RC$ to a steady-state, randomly fluctuating voltage level. In accordance with Boltzmann's energy equipartition theorem, this will result in $kT/2$ mean energy $E_C$ in the capacitor (because the capacitor is an energy storage element with one thermodynamical degree of freedom [14]):

$$E_C = \frac{1}{2}kT = \frac{1}{2}C\langle U^2(t)\rangle = \frac{1}{2C}\langle Q^2(t)\rangle \quad , \tag{1}$$

where $k=1.38*10^{-23}$ ($J/K$) is the Boltzmann constant; $R$ is the resistance, $C$ is the capacitance; $U(t)$ is the instantaneous voltage amplitude on the capacitor; $Q(t)$ is the corresponding instantaneous charge on the capacitor plates; and the symbol $\langle ...\rangle$ represents either time average in a single system, or ensemble average (at a given time instant) in a large number of identical independent systems. The capacitor plates are charged to oppositely with equal absolute instantaneous charge value, which, in accordance with Equation 1, is given as:

$$\langle Q^2(t)\rangle = kTC \tag{2}$$

In accordance with the Coulomb law, opposite charges will attract each other thus allowing one of the capacitor plates to move toward the other one, which will result in a positive mechanical work executed by the moving plate. Equation 2 indicates that the force is proportional to the absolute temperature therefore we can utilize this fact to construct electrical heat engines directly utilizing the fluctuation-dissipation theorem. Having a purely electrical heat engine has certain advantages. For example, no steams, gases, liquids, photons, phase transitions are used, no exhaust/pollution is generated, and we can use *voltage*-controlled switches (*free from any mechanical motion*) instead of valves or levers to control the engine.

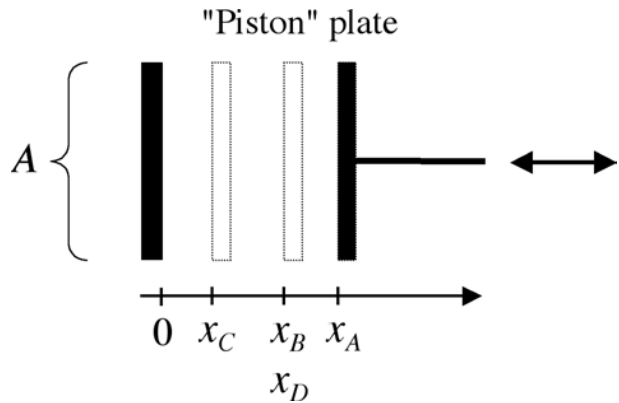

**Figure 1.** A parallel-plate capacitor's right plate (square plate with size *A*) is moving along the x axis and functioning as a "piston" driven by the resultant of the Coulomb force and external force. The left plate is fixed with its surface positioned at coordinate $x=0$. The other labels on the *x* coordinate and the pistons with dashed lines indicate the positions of the piston surface at the boundaries of different strokes in a 4-stroke cycle.

In the engines to be introduced below, the moving capacitor plate has the role of the piston and the unidirectional motion of the piston during a step within a full cycle will be called a stroke, see Figure 1. We will show 4-stroke and 2-stroke cycles. Positive work is provided during strokes when the capacitor is contracting and negative work during expansion; situations opposite to the ones in usual heat engines.

## 2.2 Block diagram of thermal noise engines and the energy price of switch control

In the analysis of the engines below, we focus on the physical principle and the fundamental properties of the idealized schemes. We do not discuss its engineering design or its mechanical feasibility with today's nanomechanics technology. However, concerning physical feasibility, we have to analyze the energy dissipation of driving the ensemble of synchronized switches because of the large number of switches and the small work a single cylinder offers during a single cycle.

The block diagram of the generic thermal noise engine is shown in Figure 2. It is supposed that a large number *N* of parallel cylinders work in the engine to provide a large ensemble for a good statistics, deterministic operation, and negligible switching losses. The moving plates in the *N* independent capacitors are mechanically coupled to a single inertia unit, a *mechanical resonator* (not shown), which is analogous to the *flywheel* as in classical heat engines. It has sufficiently large inertia to keep a nearly constant angular frequency of the sinusoidal oscillation of the piston during the strokes with positive or negative work. During contraction, the cylinders contribute energy to the flywheel's kinetic energy and during expansion, the flywheel provides necessary kinetic energy for the work on the capacitor plate against the Coulomb attraction. The electrical generator is a necessity because some electricity is needed to drive the electronic switches. Note, in a practical design, the generated electricity may also be the energy output of the engine due to the high frequency of engine motion and piezoelectric generators are one of the candidates due to their high (>95%) efficiency [13].

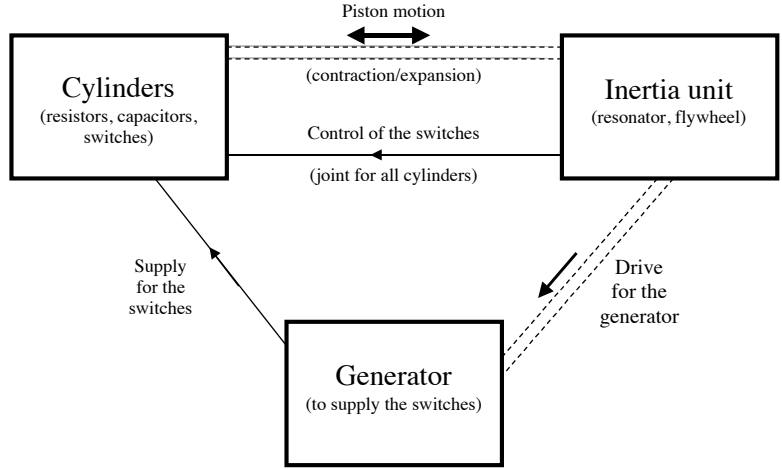

**Figure 2.** Block diagram of the generic thermal noise engine. $N \gg 1$ cylinders are working parallel in a synchronous way and driving the inertia unit, which is a flywheel in usual heat engines. However, in thermal noise engines, it would probably be a (nano-) mechanical resonator integrated on the chip.

These engines are similar to the Maxwell demon [14] (and its newer versions such as the engines of Szilard [15], and Gea-Banacloche and Leff [16]) in one aspect: in a single "cylinder" of the engine, during a single cycle, the available work is of the order of $kT$. However, in all the other aspects, the thermal noise engine is very different from these demons because neither measurements nor related decisions and separate actions about the different cylinders are needed during the work cycle. In this way, the Carnot efficiency limit can be approached because, due to the synchronous operation of the cylinders and their large number, the energy requirement for driving the switches can be made a negligible, see below.

Due to the "space/time periodicity" of this system, that is, the parallel, synchronous and periodic operation of cylinders, and their large number $N$, the energy dissipation of switches is a negligible loss for the following reasons. First we show that the energy requirement for switching operations with controlled error probability is negligible. For the synchronized switches of the $N$ cylinders, field effect transistors (MOSFETs, single-electron transistors, etc) can be used with parallel-connected gates where each drain/source electrode pair acts as a corresponding synchronized switch in one of the cylinders. The gate are wired together and their capacitance form a single capacitor (see Figure 3) with $kT/2$ total thermal energy, independently of the number $N$ of synchronized switches and parallel gates. During the on/off operation of the switch, the control voltage on the capacitor must be alternated between two different levels, which implies an energy difference $E_s$ necessary to drive the switch. To provide such an operation with error probability $\varepsilon$, the minimal energy is $E_s = -kT \ln\left(\frac{\sqrt{3}}{2}\varepsilon\right)$ [17], which, for $\varepsilon = 0.1$, 0.01, and 0.001 yields $E_s \approx 2.5kT$, $4.8kT$ and $7.1kT$, respectively. Because this energy requirement is independent of $N$, in the case of sufficiently large number of cylinders, it is a negligible loss compared to the total work offered by the $N$ cylinders. On the other hand, increasing *N* at fixed energy in the resultant capacitor implies a decreasing control voltage on the capacitor. That means that beyond a certain *N* value, the voltage difference will not be enough to control the switches (to open or close the transistor switches). The voltage cannot be allowed to decrease beyond this limit. This requirement implies that, for larger *N* values, the energy loss will grow proportionally with *N*. However, the periodic operation in time can be utilized to decrease this energy loss: the joint gate electrodes can be connected parallel to an inductor to form an *LC* resonator with the same resonance frequency $1/\sqrt{2\pi LC}$ as that of the cyclic motion of the engine, se Figure 3. If the quality factor of the *LC* circuit is *Q* then the energy dissipation during a cycle will decrease by *Q*-fold. Supposing *Q*=100, for $\varepsilon = 0.1, 0.01$, and $0.001$ and using the energy limits given above, we get $E_s \approx 0.025 kT$, $0.048 kT$ and $0.071 kT$, respectively, as the lower limits of switching energy loss. Note, in a practical design an inductor may replaced by a piezo resonator and it maybe the same resonator which is used as flywheel.

In conclusion, in the large *N* and *Q* limit, the energy loss due to the operation of switches is negligible.

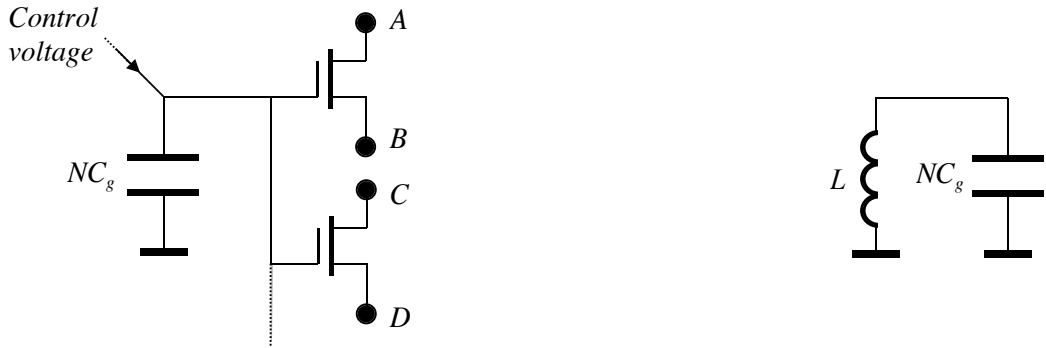

**Figure 3.** Running the *N* switches in a synchronized and time-periodic way (space/time periodicity) can make the energy requirement of driving the switches negligible compared to the energy output of the engine.

*2.3 Practical assumptions and their implications*

These assumption hold for all the different engines and cycles to be shown below. The duration of a single stroke $\tau_w$ and characteristic on/off transition time $\tau_s$ of switches must satisfy:

$$\tau_s \ll RC_m < RC_M \ll \tau_w \qquad (3)$$

where *R* is the driving resistance, and $C_m$ and $C_M$ are the minimum and maximum values of the capacitance during the cycle, respectively. The right inequality guarantees that the energy in the capacitor is in thermal equilibrium (isothermal operation) when the resistor is connected to the capacitor and the left inequality guarantees that the switching is so abrupt that it does not interfere with the functioning of the engine. For the sake of simplicity (but without limiting generality) we suppose to have parallel-plate capacitors (Figure 1) with square-shaped plates and suppose that the distance $x$ between the plates is always much less than the size $A$ of the plates:

$$x \ll A \tag{4}$$

In this case, the capacitance and the attractive force $F$ between the capacitor plates are:

$$C = \varepsilon_0 A^2 / x \quad , \tag{5}$$

$$F = Q^2 / 2\varepsilon_0 A^2 \tag{6}$$

where $\varepsilon_0$ is the dielectric constant of vacuum.

### *2.4 On the size of the cylinders*

Since the work output during a cycle does not depend on the size of cylinders, but on their numbers (see the details in Section 3), the cylinder size should be as small as possible. The small size allows $N$ to be large with high cycle frequency, therefore it provides a high power output. A lower limit of size is implied by the mean free path of charge carriers, which is in the order of 10 nanometers.

### **3. The standard thermal noise engine (non-coherent cylinder states)**

### *3.1 The cylinders*

The *standard thermal noise engine* is the simplest version of these engines. The schematic of the "cylinders" (two of the $N$ cylinders) is shown in Figure 4. Each cylinder has 4 elements: a "hot" resistor thermally contacted to the warm-source; a "cold" resistor thermally contacted to the cold-sink; the capacitor with the piston plate; and a 3-stage switch that is synchronized with the switches of the other cylinders and the flywheel (or mechanical resonator). The 3 stages of the switch are:

*H (Hot)*: The capacitor is connected to the "hot" resistor with temperature $T_H$ for an isothermal stroke at temperature $T_H$.

0: The capacitor is open-ended for adiabatic stroke.

*C (Cold)*: The capacitor is connected to the "cold" resistor with temperature $T_C$ for an isothermal stroke at temperature $T_C$.

The mean force provided by the pistons in the $N$ parallel and synchronized cylinders is:

$$\langle F \rangle = N \frac{\langle Q^2 \rangle}{2\varepsilon_0 A^2} = N \frac{C^2 \langle U^2 \rangle}{2\varepsilon_0 A^2} = N \frac{kT}{2} \frac{C}{\varepsilon_0 A^2} = N \frac{kT}{2} \frac{1}{x}. \qquad (7)$$

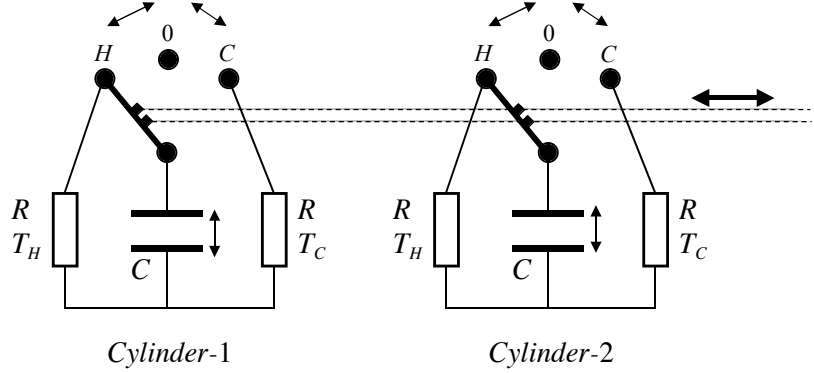

**Figure 4.** Cylinders of the standard thermal noise driven engine. Only two of the $N$ identical and parallel cylinders are shown. The double dashed lines symbolize synchronous switch operation via a mechanical connection to the flywheel section (not shown). Similar synchronous mechanics connects the flywheel with the capacitor plates in the cylinders. The expansion and contraction of the capacitors is done in synchronized and continuous fashion while the switching steps are done in a synchronized and abrupt way at the beginning of each stroke.

## *3.2 Four-stroke operation (Carnot-cycle)*

*The four-stroke operation* is a Carnot-cycle [18], see its entropy-temperature graph in Figure 5. The cycle of the cylinders begins at point $A$, with switch in the position $H$ (*Hot*, the temperature is $T_H$), and the piston is at position $x_A$. The 4 strokes are as follows:

1. Stroke $A \Rightarrow B$: Isothermal contraction, with switch in the position *Hot*, to $x_B = x_A / \alpha$ (where $\alpha > 1$). The engine will produce a positive work:

$$W_{AB} = T_H(S_B - S_A) = -\int_{x_A}^{x_B} N \frac{kT_H}{2x} dx = N \frac{kT_H}{2} \ln\left(\frac{x_A}{x_B}\right) = N \frac{kT_H}{2} \ln \alpha \qquad (8)$$

This process is isothermal thus the required heat $Q_{AB}$ from the warm-source is equal to the work produced:

$$Q_{AB} = W_{AB} = \frac{Nk}{2} T_H \ln \alpha \qquad (9)$$

2. Stroke $B \Rightarrow C$: Adiabatic, reversible ($S_C = S_B$) contraction, with switch in the position 0, to $x_C = x_B / \beta$, where $\beta$ (>1) is chosen properly to decrease the equivalent temperature of the system of $N$ capacitors from $T_H$ to $T_C$ and their energy from $NkT_H/2$ to $NkT_L/2$. The engine will produce a positive work $W_{BC} = Nk(T_H - T_L)/2$.

3. Stroke $C \Rightarrow D$: Isothermal expansion, with switch in the position $C$ (*Cold*, the temperature is $T_C$), to $x_D = x_C\,\alpha$. The work of the engine will be negative (the engine must use some of the flywheel's kinetic energy for expansion):

$$W_{CD} = T_C(S_D - S_C) = -\int_{X_C}^{X_D} N\frac{kT_C}{2x}dx = N\frac{kT_C}{2}\ln\left(\frac{x_D}{x_C}\right) = -N\frac{kT_C}{2}\ln\alpha \tag{10}$$

This process is isothermal thus the engine converts this work into heat $Q_{CD}$ which is dissipated in the resistors $R_C$, and absorbed by the cold-sink:

$$Q_{CD} = -W_{CD} = \frac{Nk}{2}T_C \ln\alpha \tag{11}$$

4. Stroke $D \Rightarrow A$: Adiabatic, reversible ($S_A = S_D$) expansion, with switch in the position 0, to $x_A = x_D\,\beta$. This is the exactly reversed operation of the second stroke thus the equivalent temperature of the system of $N$ capacitors will increase to $T_C$ from $T_H$. The work of the engine will be negative $W_{DA} = -W_{BC}$. Thus the total work of the engine during the 2nd and the 4th strokes is zero because of the reversed operations.

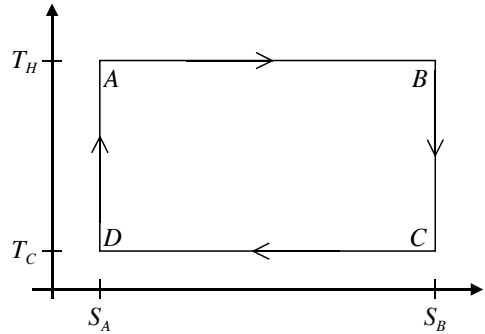

**Figure 5.** Entropy-temperature graph of the standard engine in Carnot-cycle operation mode (four-strokes). *AB* stroke: isothermal contraction (executing positive work); *BC* stroke: adiabatic contraction (executing positive work); *CD* stroke: isothermal expansion (negative work); *DA* stroke: adiabatic expansion (negative work exactly compensating the positive work during *BC*).

In conclusion, the total work produced by of the 4-stroke operation during a full cycle is:

$$W_{tot} = W_{AB} + W_{BC} + W_{CD} + W_{DA} = \frac{Nk}{2}(T_H - T_C)\ln\alpha \quad . \tag{12}$$

The total input heat consumed from the warm-source is:

$$Q_H = \frac{Nk}{2} T_H \ln \alpha \;, \tag{13}$$

and the waste heat, which is transmitted to the cold-sink, is:

$$Q_C = Q_H \frac{T_C}{T_H} \tag{14}$$

The energy efficiency of the engine:

$$\eta_{clas} = \frac{W_{tot}}{Q_H} = 1 - \frac{T_C}{T_H} \tag{15}$$

which is exactly the efficiency of the idealistic Carnot-engine [18]; the highest possible efficiency of any heat engine without violating the second law of thermodynamics.

### *3.3 Two-stroke operation*

The 2nd stroke, the adiabatic contraction, and its reversed operation, the 4th stroke, can be left out from the cycle. The switch is alternating between the *H* and *C* positions and the 0 position is not used. After a completed stroke, the switch position is alternated and the capacitor system will be "thermalized" (heated or cooled, accordingly) to the new temperature with the actual *RC* relaxation time constant before the next stroke takes place. Inequality 3 guarantees that these heating and cooling processes are much shorter than the stroke durations thus they can be done at the beginning/end of the stroke without the interruption of the mechanical motion of the engine. The entropy-temperature graph is given in Figure 6. The piston positions are:

$$x_B = x_A / \alpha \;, \qquad x_{B*} = x_B \;, \qquad x_{A*} = x_A \tag{16}$$

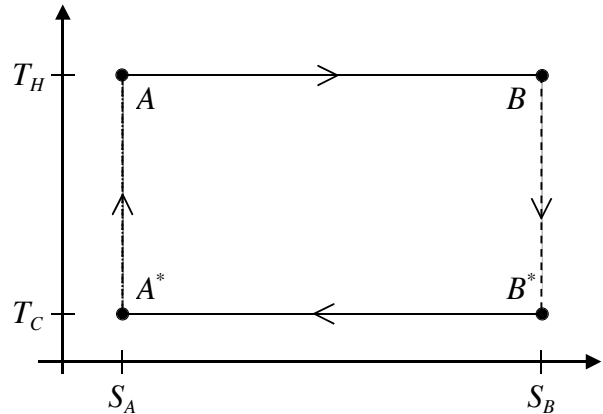

**Figure 6.** Entropy-temperature graph of the two-stroke thermal noise engine. The full cycle contains 4 steps but only 2 strokes: the isothermal constriction (*AB* stroke) and the isothermal expansion (*B\*A\** stroke). The cooling (*BB\**) and heating (*A\*A*) steps are done without piston motion and work. The solid lines represent the same strokes and piston

motion as in the 4-stroke cycle. The dashed lines represent the cooling and the heating steps without motion and work.

The heat transfer and the net work are zero during the omitted strokes, the total energy and work balance and the energy efficiency remain the same, as described by Equations 12-15.

**4. Thermal noise engines with coherent (correlated) cylinder states**

The coherent thermal noise engine is inspired by coherent quantum heat engines [7-11]. During the stroke where the work of the engine is negative (*CD* stroke in Figure 5 and *B\*A\** stroke in Figure 6), the original number *N* of thermodynamical degrees of freedom is reduced to one by switching all the capacitors parallel to form a single capacitor with *N*-times greater capacitance. Thus, in large systems, the negative work component approaches zero and the engine *seemingly* provides efficiency better than the Carnot limit and *seemingly* can work with a single temperature. For the resolution of this paradox, see the section "*The catch...*" below.

*4.1. Cylinders and operation: 2-heat-reservoir and single-heat-reservoir engines*

The schematic of the "cylinders" (two of the *N* cylinders) is shown in Figure 7. Each cylinder has 5 elements: the same as the 4 elements that applied in the standard engine (see Figure 4) and an extra "coherence" switch. In the "*Coh*" stage, the coherence switch connects the capacitor in parallel with the capacitors in the other cylinders. The coherent switches are also synchronized in all the cylinders (see Figure 7).

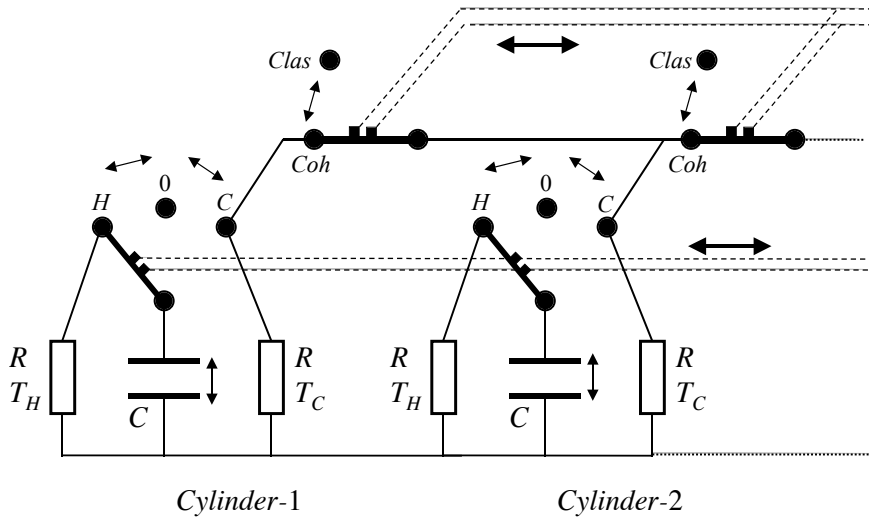

**Figure 7**. The coherent engine. During the isothermal expansion stroke (*CD* in Figure 5 and B*A* in Figure 6, when the engine consumes work from the flywheel/resonator unit) the upper switches connect all the capacitors parallel in their *Coh* (Coherent) position. Then the thermal noise in all the capacitors will be identical (coherent noises). For further clarification, see Section 4.2. During the rest of the strokes these switches are in the *Clas* (Classical) position and the operation of the engine remains the same as in the standard engine.

Thus, in the "$Coh$" stage, the number of thermodynamical degrees of freedom in the system is reduced from the original number $N$ to 1, and that yields the reduction of force and the negative work by the same factor. Therefore, in the coherent engine, the negative work, instead of Equation 11, becomes:

$$W_{CD,coh} = -Q_{CD} = -\int_{x_C}^{x_D} \frac{kT_C}{2x} dx = \frac{kT_C}{2}\ln\left(\frac{x_D}{x_C}\right) = -\frac{kT_C}{2}\ln\alpha \quad . \tag{17}$$

The total work of the coherent engine is greater than in Equation 12:

$$W_{tot,coh} = W_{AB} + W_{BC} + W_{CD} + W_{DA} = \frac{1}{2}k(NT_H - T_C)\ln\alpha \quad , \tag{18}$$

and, because the input remains the same as described by Equation 13, the energy efficiency appears to be greater than the Carnot-efficiency and, for large $N$, it approaches 100% which contradicts to the Second Law of Thermodynamics:

$$\eta_{coh} = \frac{W_{tot,coh}}{Q_H} = \frac{k(NT_H - T_C)}{NkT_H} = 1 - \frac{T_C}{NT_H} \tag{19}$$

Similar statements had emerged about quantum-heat-engines before the issues got clarified [7-11]. But the similarity between coherent thermal noise and quantum engines goes further: the above equations clearly indicate that the coherent engine can produce useful work and has seemingly near 100% efficiency even when $T_H = T_C$, which are a direct violations of the Second Law:

$$W_{tot,coh,T_H} = \frac{1}{2}kT_H(N-1)\ln\alpha \qquad \text{and} \tag{20}$$

$$\eta_{coh,T_H} = \frac{W_{tot,coh}}{Q_H} = \frac{kT_H(N-1)}{NkT_H} = 1 - \frac{1}{N} \tag{21}$$

Accordingly, a simplified coherent engine, with seemingly a single heat reservoir, can be constructed, see Figure 8, that is also described by Equations 20-21.

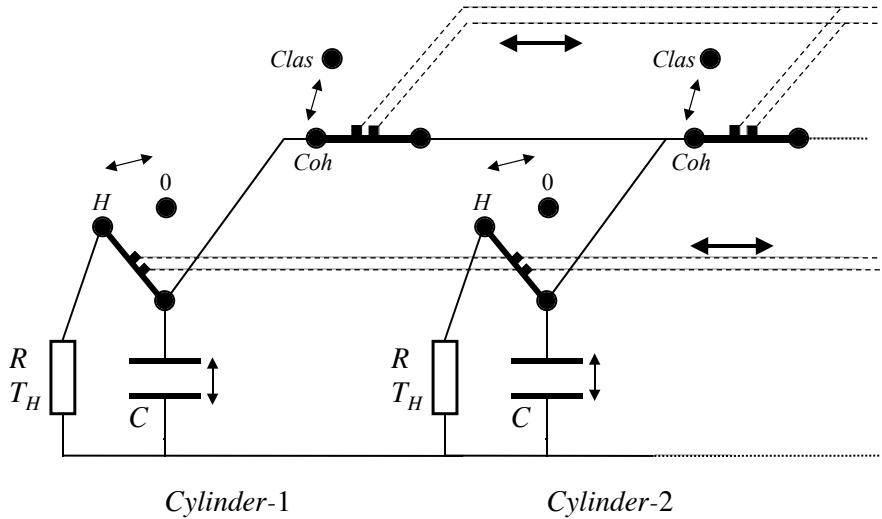

*Cylinder*-1                     *Cylinder*-2

**Figure 8.** Coherent engine with (seemingly) a single heat reservoir. During the isothermal expansion stroke (*CD* in Figure 5 and B*A* in Figure 6, when the engine consumes work from the flywheel/resonator unit) the upper switches connect all the capacitors parallel in their *Coh* (Coherent) position and the lower switch is in the *H* position.

Figure 9 shows the temperature-entropy graph of the single-heat-reservoir-two-stroke engine with coherent cylinders during the stroke with negative work. There is positive work during the *AB* stroke. Then, when the coherence is established, the system jumps to point *B\** and the second stroke *B\*A* needs a reduced work from the inertia system and, accordingly, a reduced entropy change to restore the initial condition *A* .

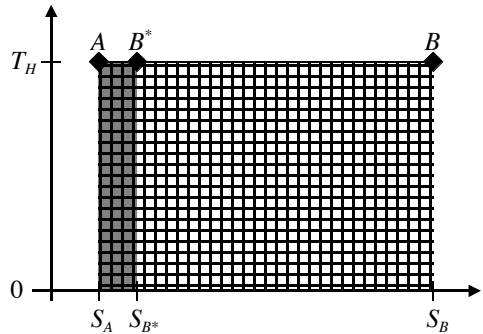

**Figure 9.** The entropy-temperature graph of the two-strokes engine with a single heath reservoir and coherent (correlated) cylinders, cf with the non-coherent version in Figure 6. The first stroke goes from *A* to *B* and executes the work represented by the area below the *AB* section (grid pattern). Then, at the introduction of coherence, the entropy jumps from *B* to *B\** and the second stroke goes from *B\** to *A* and it requires the smaller amount of work represented by the area below the *B\*A* section (gray filling).

## 4.2 The catch: where does the entropy and the heat go?

To solve the paradox of the seeming violation of the Second Law by the coherent engines, first the following question must be answered: *Where does the entropy (see Figure 6) and the energy go when the coherence is established switched on*? Due to Equation 1, at point *B*, the *rms* charge in a

single capacitor is:

$$\sqrt{\langle Q_i^2 \rangle_N} = \sqrt{kT_H C} \tag{22}$$

where the average is taken over the ensemble of the $N$ capacitors. At the moment when the coherence is switched on and the system jumps from $B$ to point $B^*$, due to $\langle Q_i \rangle_N = 0$ and the charge conservation the central limit theorem applies for the effective charge at $B^*$ in the resultant single capacitor with capacitance $C_{B^*} = NC$:

$$Q_{B^*} = \sqrt{NkT_H C} \quad , \tag{23}$$

and

$$E_{B^*} = \frac{Q_{B^*}^2}{2C_{B^*}} = \frac{NkT_H C}{2NC} = \frac{kT_H}{2} \ . \tag{24}$$

That means, $(N-1)kT_H/2$ energy disappeared from the system and, because the same result holds even when no resistors are connected to the capacitors, the disappeared energy does not go back to the heat-source. The solution is well known, namely, that the energy at charge equilibration in parallel capacitors is dissipated in the stray resistance of the connecting wires (and switches) and that the energy loss is independent of the actual stray resistance values. This fact indicates that it is necessary to install a heat reservoir (with temperature $T_{S,W}$) around the wires and the coherence switches otherwise they will overheat. This originally hidden heat reservoir and the stray resistances $R_{S,W}$ will also act as thermal noise generators and cause thermal noise loop-currents through the capacitors, in accordance with the fluctuation-dissipation theorem, see Figure 10. The system of parallel capacitors will be thermalized to temperature temperature $T_{S,W}$ in a very short time, in the order of $R_{S,W}C$ which is the shortest electrical time constant in the system.

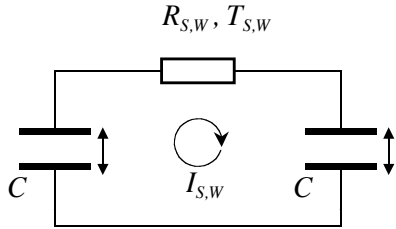

**Figure 10.** The loop currents generated by the thermal noise of stray resistance of switches and wires.

Thus, equations 17-21 and the enhanced properties of the coherent engines are valid only with idealistic switches and wires that have zero thermal noise. This condition is satisfied only at absolute zero temperature, $T_{S,W} = 0$ Kelvin. This is an indication that, when the coherence is introduced with the assumed properties, a new, hidden thermal reservoir around the wire connections is implicitly introduced and that takes over the role of the cold sink during the coherent stroke. In this way, Carnot's work and efficiency are restored and the Second Law is satisfied. The relevance of this explanation for the case of coherent quantum-heat engines is also obvious: the thermal reservoir (or possibly a heat pump) around the coherent medium is necessary to remove the exhaust heat and this new heat reservoir will play the role of the cold sink during the coherent stroke and preserves the Second Law and the Carnot properties.

The coherent engine turns out to be no better than the standard engine. However, in certain practical designs the connecting wires and switches can be used as cold resistors and connected to the cold sink in a coherent engine.

**5. Estimation of work and efficiency at an idealistic example**

Assume a 1 square inch (2.5x2.5cm) chip on which thermal noise engines of 25 nanometer effective size are integrated, thus $N=10^{12}$. The Carnot efficiency can be approached if the quality factors of the LC and mechanical resonators are sufficiently large which is a reasonable assumption. At this length scale, it is practical to suppose 10 *GHz* mechanical resonator frequency which is the cycle frequency of the engines. The noise bandwidth of electronic element of this size is >500 *GHz* thus Inequality 3 is satisfied. Let us suppose we want to harvest energy from a 10 ºC temperature gradient at room temperature: the cold sink is at 25 ºC and the warm source at 35 ºC. The work output of the idealistic engine would be 0.48 *Watt*.

**Acknowledgement**

Comments from Sunil Khatri, Andreas Klappenecker, Julio Gea-Banacloche, Ferdinand Peper and Hank Walker are appreciated.

**References**

[1] J.B. Johnson, "Thermal agitation of electricity in conductors", *Nature* **119** (1927) 50-51.
[2] H. Nyquist, "Thermal agitation of electric charge in conductors", *Physical Review* **32** (1928) 110-113.
[3] M. Born, W. Heisenberg, P. Jordan, "Quantum mechanics II", *Zeitschrift für Physik* **35** (1926) 557-615.
[4] D. Jason Palmer, "Noise encryption keeps spooks out of the loop", *New Scientist*, issue 2605 (23 May 2007), 32; http://www.newscientist.com/article/mg19426055.300-noise-keeps-spooks-out-of-the-loop.html
[5] Adrian Cho, "Simple noise may stymie spies without quantum weirdness", *Science* **309** (2005) 2148; http://www.ece.tamu.edu/~noise/news_files/science_secure.pdf

[6] Justin Mullins, "Breaking the noise barrier", New Scientist, issue 2780 (29 September 2010); http://www.newscientist.com/article/mg20827801.500-breaking-the-noise-barrier.html?full=true

[7] A.E. Allahverdyan, T.M. Nieuwenhuizen, "Extraction of work from a single thermal bath in the quantum regime", *Physical Rev. Lett.* **85** (2000) 1799-1802.

[8] M.O. Scully, M.S. Zubairy, G.S. Agarwal, H. Walther, Extracting work from a single heat bath via vanishing quantum coherence, *Science* **299** (2003) 862–864.

[9] D. Jou, J. Casas-Vazquez, "About some current frontiers of the second law", *J. Non-Equilib. Thermodyn.* **29** (2004) pp. 345–357.

[10] T. Zhang, L.F. Cai, P.X. Chen, C.Z. Li, "The Second Law of Thermodynamics in a Quantum Heat Engine Model", *Commun. Theor. Phys.* **45** (2006) 417–420.

[11] J. Arnaud, L. Chusseau, F. Philippe, "Mechanical equivalent of quantum heat engines", *Physical Rev. E* **77** (2008) 061102.

[12] H. B. Callen and T. A. Welton, "Irreversibility and Generalized Noise", *Physical Rev.* **83** (1951) 34-40.

[13] Evans, K. A., "Power generation efficiency of piezoelectric crystals", Ph.D. Thesis Georgetown Univ., Washington, DC., 1984.

[14] Maxwell J. C., Theory of Heat (Longmans, Green and Co., London) 1871.

[15] Szilard L., Z. Phys., 53 (1929) 840-856.

[16] J. Gea-Banacloche, H.S. Leff, "Quantum, version of the Szilard one-atom engine and the cost of raising energy barriers", *Fluctuation and Noise Lett.*, **5** (2005) C39–C47

[17] L.B. Kish, "Moore's Law and the Energy Requirement of Computing versus Performance", *IEE Proc. - Circ. Dev. Syst.* **151** (2004) 190-194.

[18] Sadi Carnot (and R.H. Thurston, editor/translator) (1890), "Reflections on the Motive Power of Heat and on Machines Fitted to Develop This Power", J. Wiley&Sons., New York, 1890; online: http://www.archive.org/details/reflectionsonmot00carnrich .